\documentclass[aps,prl,reprint,superscriptaddress,showpacs]{revtex4-1} 
\usepackage{graphicx}  
\usepackage{dcolumn}   
\usepackage{amssymb}
\usepackage{physics}
\usepackage{natbib}
\usepackage{color}

\begin{document}

\title{Superconducting Gatemon Qubit based on a \\ Proximitized Two-Dimensional Electron Gas}
\author{L.~Casparis}
\altaffiliation{Contributed equally to this work}
\affiliation{Center for Quantum Devices, Station Q Copenhagen, Niels Bohr Institute, University of Copenhagen, Copenhagen, Denmark}
\author{M.~R.~Connolly}
\altaffiliation{Contributed equally to this work}
\affiliation{Center for Quantum Devices, Station Q Copenhagen, Niels Bohr Institute, University of Copenhagen, Copenhagen, Denmark}
\author{M.~Kjaergaard}
\altaffiliation[Present address: ]{Research Laboratory of Electronics, Massachusetts Institute of Technology, Cambridge, MA 02139, USA}
\affiliation{Center for Quantum Devices, Station Q Copenhagen, Niels Bohr Institute, University of Copenhagen, Copenhagen, Denmark}
\author{N.~J.~Pearson}
\affiliation{Center for Quantum Devices, Station Q Copenhagen, Niels Bohr Institute, University of Copenhagen, Copenhagen, Denmark}
\affiliation{Theoretische Physik, ETH Z{\"u}rich, Z{\"u}rich, Switzerland}
\author{A.~Kringh\o{}j}
\affiliation{Center for Quantum Devices, Station Q Copenhagen, Niels Bohr Institute, University of Copenhagen, Copenhagen, Denmark}
\author{T.~W.~Larsen}
\affiliation{Center for Quantum Devices, Station Q Copenhagen, Niels Bohr Institute, University of Copenhagen, Copenhagen, Denmark}
\author{F.~Kuemmeth}
\affiliation{Center for Quantum Devices, Station Q Copenhagen, Niels Bohr Institute, University of Copenhagen, Copenhagen, Denmark}
\author{T.~Wang}
\affiliation{Department of Physics and Astronomy, Purdue University, West Lafayette, Indiana 47907, USA}
\affiliation{Station Q Purdue, and Birck Nanotechnology Center, Purdue University, West Lafayette, Indiana 47907, USA}
\author{C.~Thomas}
\affiliation{Department of Physics and Astronomy, Purdue University, West Lafayette, Indiana 47907, USA}
\affiliation{Station Q Purdue, and Birck Nanotechnology Center, Purdue University, West Lafayette, Indiana 47907, USA}
\author{S.~Gronin}
\affiliation{Station Q Purdue, and Birck Nanotechnology Center, Purdue University, West Lafayette, Indiana 47907, USA}
\author{G.~C.~Gardner}
\affiliation{Station Q Purdue, and Birck Nanotechnology Center, Purdue University, West Lafayette, Indiana 47907, USA}
\author{M.~J.~Manfra}
\affiliation{Department of Physics and Astronomy, Purdue University, West Lafayette, Indiana 47907, USA}
\affiliation{Station Q Purdue, and Birck Nanotechnology Center, Purdue University, West Lafayette, Indiana 47907, USA}
\affiliation{School of Materials Engineering, Purdue University, West Lafayette, Indiana 47907, USA}
\affiliation{School of Electrical and Computer Engineering, Purdue University, West Lafayette, Indiana 47907, USA}
\author{C.~M.~Marcus}
\affiliation{Center for Quantum Devices, Station Q Copenhagen, Niels Bohr Institute, University of Copenhagen, Copenhagen, Denmark}
\author{K.~D.~Petersson}
\email{karl.petersson@nbi.ku.dk}
\affiliation{Center for Quantum Devices, Station Q Copenhagen, Niels Bohr Institute, University of Copenhagen, Copenhagen, Denmark}

\maketitle

%

\textbf{The coherent tunnelling of Cooper pairs across Josephson junctions (JJs) generates a nonlinear inductance that is used extensively in quantum information processors based on superconducting circuits, from setting qubit transition frequencies \cite{koch_2007} and interqubit coupling strengths \cite{chen_2014}, to the gain of parametric amplifiers \cite{castellanos-beltran_2007} for quantum-limited readout. The inductance is either set by tailoring the metal-oxide dimensions of single JJs, or magnetically tuned by parallelizing multiple JJs in superconducting quantum interference devices (SQUIDs) with local current-biased flux lines. JJs based on superconductor-semiconductor hybrids represent a tantalizing all-electric alternative. The \textit{gatemon} is a recently developed transmon variant which employs locally gated nanowire (NW) superconductor-semiconductor JJs for qubit control~\cite{larsen_2015,delange_2015}. Here, we go beyond proof-of-concept and demonstrate that semiconducting channels etched from a wafer-scale two-dimensional electron gas (2DEG) are a suitable platform for building a scalable gatemon-based quantum computer. We show 2DEG gatemons meet the requirements~\cite{divincenzo_2000} by performing voltage-controlled single qubit rotations and two-qubit swap operations. We measure qubit coherence times up to $\sim$~2 $\mu$s, limited by dielectric loss in the 2DEG host substrate.}

\begin{figure}[!b]\vspace{-4mm}
    \includegraphics[width=1\columnwidth]{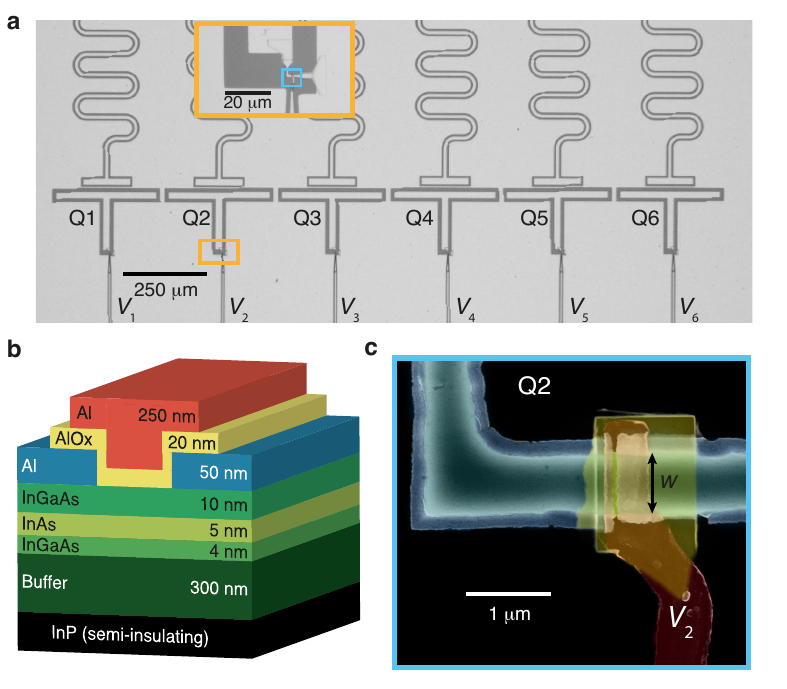}
    \caption{\textbf{2DEG gatemon.} \textbf{a,} Optical micrograph of a six qubit device. The 2DEG JJ is shunted by the T-shaped island to the surrounding ground plane and coupled to individual readout cavities. The gate voltage $V$\textsubscript{j} changes the qubit frequency of Qj. \textbf{b,} Schematic of the wafer stack. \textbf{c,} False coloured scanning electron micrograph of the gate controlled 2DEG JJ of width $w$.}
    \label{device}
\end{figure}

Figure~\ref{device}(a) shows an optical micrograph of a typical device hosting six 2DEG gatemon qubits. Each gatemon comprises an Al island shunted to the ground plane via a 2DEG JJ and capacitively-coupled to a serpentine-shaped coplanar waveguide cavity. The self capacitance $C$ of the island together with the nonlinear inductance of the JJ creates an anharmonic potential for plasmon oscillations across the JJ. The ground $|0\rangle$ and excited $|1\rangle$ states of the qubit correspond to the lowest two harmonic oscillator states, which in the transmon limit ($E_J \gg E_C$) are separated in energy by a transition frequency $f_Q \approx \sqrt{8E_CE_J}/h$, where $E_C = e^2/2C$ is the charging energy, and $E_J$ is the Josephson energy ~\cite{koch_2007, clarke_2008}.
 
\begin{figure*}[t]
    \centering
        \includegraphics[width=2\columnwidth]{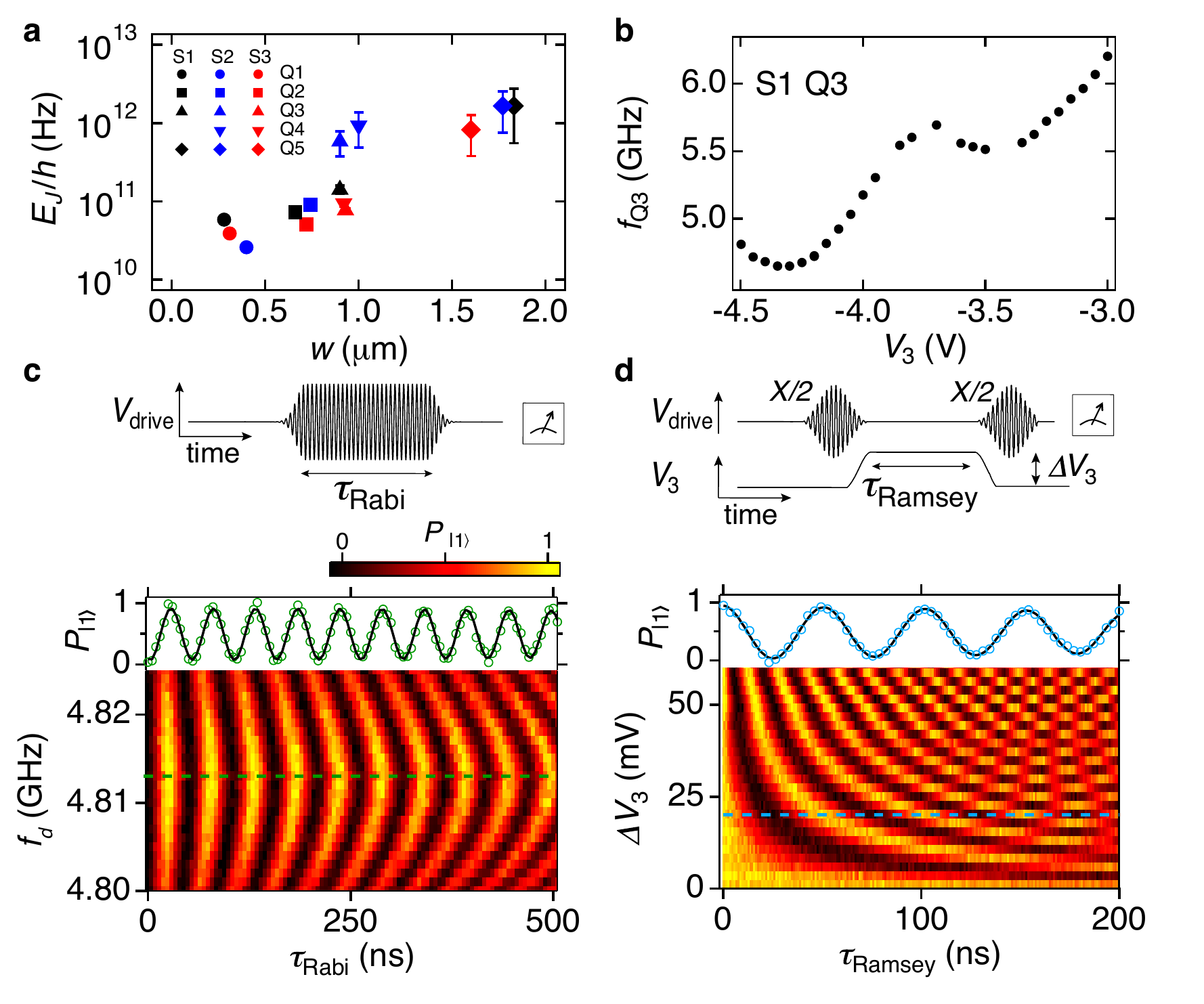}\vspace{-4mm}
    \caption{\textbf{Coherent qubit manipulation.} \textbf{a,} Overview of $E_J$ as a function of $w$ for three devices S1, S2 and S3 at zero gate voltage. \textbf{b,} Frequency of S1 Q3 as a function of gate voltage. \textbf{c,} Coherent Rabi oscillations performed at $V_3$~=~-4.5~V by applying the microwave pulse sequence shown in the upper panel. The main panel shows qubit oscillations as a function of qubit drive frequency and $\tau_{\rm{Rabi}}$ with the inset showing a cut at the resonance frequency. The solid line is a fit to a Gaussian damped sinusoid. \textbf{d,} Coherent qubit rotations around the $Z$-axis. The qubit is positioned on the equator with an X/2 pulse followed by a gate pulse with amplitude $\Delta V_3$ and duration $\tau_{\rm{Ramsey}}$ and finally rotated back by an X/2 pulse, upper panel. The main panel shows the coherent $Z$ oscillation as a function $\Delta V_3$ and $\tau_{\rm{Ramsey}}$ with the inset showing a cut at $\Delta V_3$~=~20~mV.}
    \label{coh}
\end{figure*}

Fixed frequency transmons, employing single metal-oxide JJs, benefit from longer coherence times, but at the cost of slow ($\sim$~150 ns) two-qubit gate operation times \cite{sheldon_2016} and frequency crowding \cite{hutchings_2017}. Frequency tunable qubits allow faster two-qubit gates, but flux noise of $\sim$~1 $\mu \Phi_0/\sqrt{Hz}$ in SQUIDs limits phase coherence to $T_2^* \sim$~5~$\mu$s~\cite{kelly_2015}. Moreover, the milliampere currents used to control the flux in the SQUIDs place additional demands on cooling power and complicate the integration with 3D architectures. Qubit-qubit and qubit-coupler crosstalk can also exceed 4$\%$ and requires calibration protocols that scale with the number of qubits~\cite{Neill_2017}. In superconductor-semiconductor JJs, $E_J$ can be controlled by local, capacitively-coupled gates \cite{delange_2015,larsen_2015,kjaergaard_2017}, opening up the possibility to tune and modulate $f_Q$ without crosstalk or sensitivity to flux noise. 

In this work we realize scalable superconductor-semiconductor JJs using the 2DEG heterostructure shown schematically in Fig.~\ref{device}(b). The 2DEG is formed in an InAs quantum well encapsulated between InGaAs barriers. We leverage recent breakthroughs in \textit{in-situ} epitaxy of Al on III-V semiconductors~\cite{shabani_2016} to obtain a pristine high-transparency superconductor-semiconductor interface between a 50 nm-thick layer of superconducting aluminium and the 2DEG. Semi-insulating (Fe-doped) InP is used as a host substrate for the 2DEG buffer layer, which is etched away before patterning the qubit island and microwave control circuitry. Figure~\ref{device}(c) shows a false coloured scanning electron micrograph of the JJ. 


First, we demonstrate that 2DEG gatemons can be fabricated deterministically with reproducible properties. We fabricated three devices (S1, S2, S3) each hosting six qubits with junction width $w$ increasing from 0.3-2.6~$\mu$m (labelled Q1-Q6). To extract $E_J$ of the as-fabricated qubits the corresponding cavity frequency is measured before any voltage is applied to the gate. Due to vacuum fluctuations in the electric field between the cavity and qubit, the cavity is Lamb-shifted from its bare resonance frequency by $\chi = g_{\rm{cav}}^2/\Delta_Q$, where $\Delta_Q/2\pi = f_c - f_Q$ is the qubit-cavity detuning and $g_{\rm{cav}}$ is the coupling strength. This expression and the directly measured $g_{\rm{cav}}/2\pi\sim$~100 MHz, together with numerical simulations for $C$ ($E_C \sim$~230 MHz), allows us to estimate $E_J$. Figure~\ref{coh}(a) plots the extracted $E_J$ as a function $w$ for all measured qubits. The data show that $E_J$ increases for wider junctions as expected for an increasing number of modes participating in Cooper pair transport~\cite{kringhoj_2017}. Such precise control of $E_J$ on a design parameter $w$ represents an important step towards engineering scalable superconductor-semiconductor quantum information processors, improving on previous realizations where $w$ was limited by the 1D character of NWs. 

Next, we show all-electric control by tuning the qubit transition frequency in Fig.~\ref{coh}(b). We operate in the transmon regime, $E_J/E_C \sim$~70-130, and read out the qubit dispersively ($g_{\rm{cav}} \ll |\Delta_Q|$)~\cite{wallraff_2004}. Using two-tone spectroscopy, we drive a single qubit (Q3) and identify its frequency as a function of gate voltage from the state-dependent push on the cavity. The frequency $f_{\rm{Q3}}(V_3)$ is monotonic over a wider voltage range than for NWs~\cite{larsen_2015,casparis_2016} and can be tuned by $\Delta f\sim$~1 GHz for 1 V applied to the gate ($V$\textsubscript{j} corresponds to the voltage applied to j-th qubit Qj). The dependence of qubit frequency on gate voltage can be optimized by changing the thickness of the dielectric layer and using 2DEGs with different field-effect mobility. 

We next demonstrate basic operations of individual qubits using time domain manipulation and readout. Phase-controlled microwave pulses with frequency $f_d$ are applied either via the cavity readout feedline or separately through the JJ top-gate. The rotation about the X-axis of the Bloch sphere is performed by applying the pulse for a time $\tau_{\rm{Rabi}}$ and reading out the state via the cavity [pulse sequence, Fig.~\ref{coh}(c)]. Plotting the probability to be in $|1\rangle$, $P_{|1\rangle}$, as a function of $\tau_{\rm{Rabi}}$ and $f_d$, reveals Rabi oscillations [Fig.~\ref{coh}(c)], characteristic of the qubit rotation. These data are used to calibrate the pulse times and amplitudes to rotate by $\pi$ and $\pi/2$ around the X-axis ($X$ and $X/2$ pulses respectively). We next show coherent accumulation of dynamical phase by controlled rotation of the qubit around the Z-axis. Fig.~\ref{coh}(d) shows the pulse sequence comprising a resonant ($f_d=f_Q$) $X/2$ pulse, a gate pulse with amplitude $\Delta V_3$ and duration $\tau_{\rm{Ramsey}}$, and a second $X/2$ pulse. When $\Delta V_3$~=~0 the qubit and drive are phase locked, so the two $X/2$ pulses rotate the qubit to the $|1\rangle$ state. With increasing $\Delta V_3$ the qubit rotates around the Z-axis relative to the drive. While further experiments are required to establish fidelities, these data establish the high degree of control afforded by electrostatically-coupled gates. 


\begin{figure}
    \centering
       \includegraphics[width=1\columnwidth]{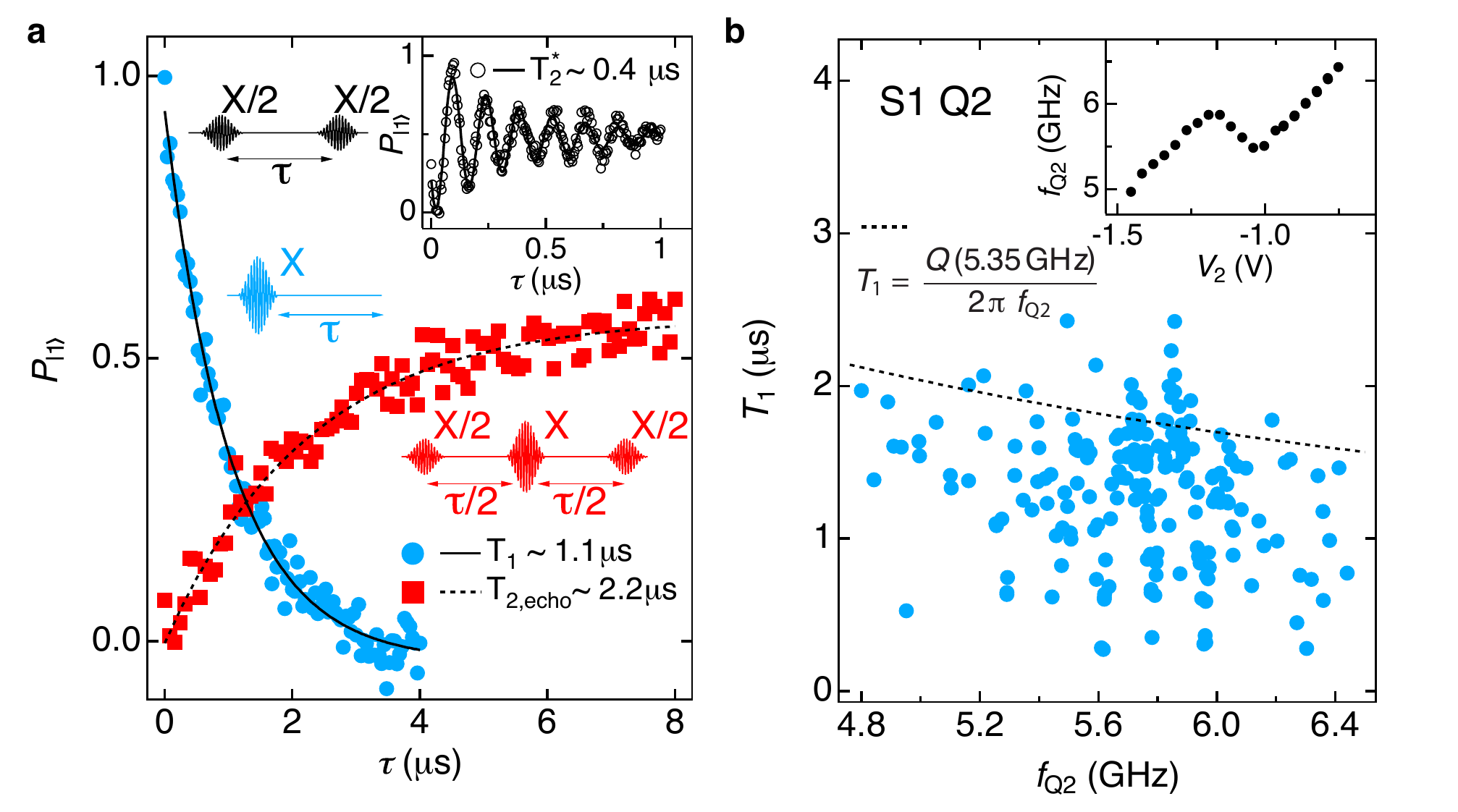}\vspace{-4mm}
    \caption{\textbf{Coherence times.} \textbf{a,} Lifetime measurement for S1 Q2 (blue) with qubit resonance frequency $f_{\rm{Q2}}\sim$~5 GHz. In red, we perform a Hahn echo experiment to determine $T_{\rm{2,echo}}$. The black lines are exponential fits. Inset: pulse sequence for dephasing ($T_{\rm{2,echo}}$) [red] and relaxation ($T_1$) [blue] measurements. The inset shows a Ramsey experiment which is performed to determine $T_2$ for Q2 with the pulse sequence shown next to the inset (black). The solid line is a fit to an exponentially damped sinusoid. \textbf{b,} Relaxation time measurements as a function of qubit frequency. The dashed line  indicates the limit on the qubit lifetime for a quality factor $Q\sim 6.4 \times 10^4$. The inset shows the spectrum for Q2.
    }
    \label{times}\vspace{-4mm}
\end{figure}

\begin{figure*}
    \centering
       \includegraphics[width=2\columnwidth]{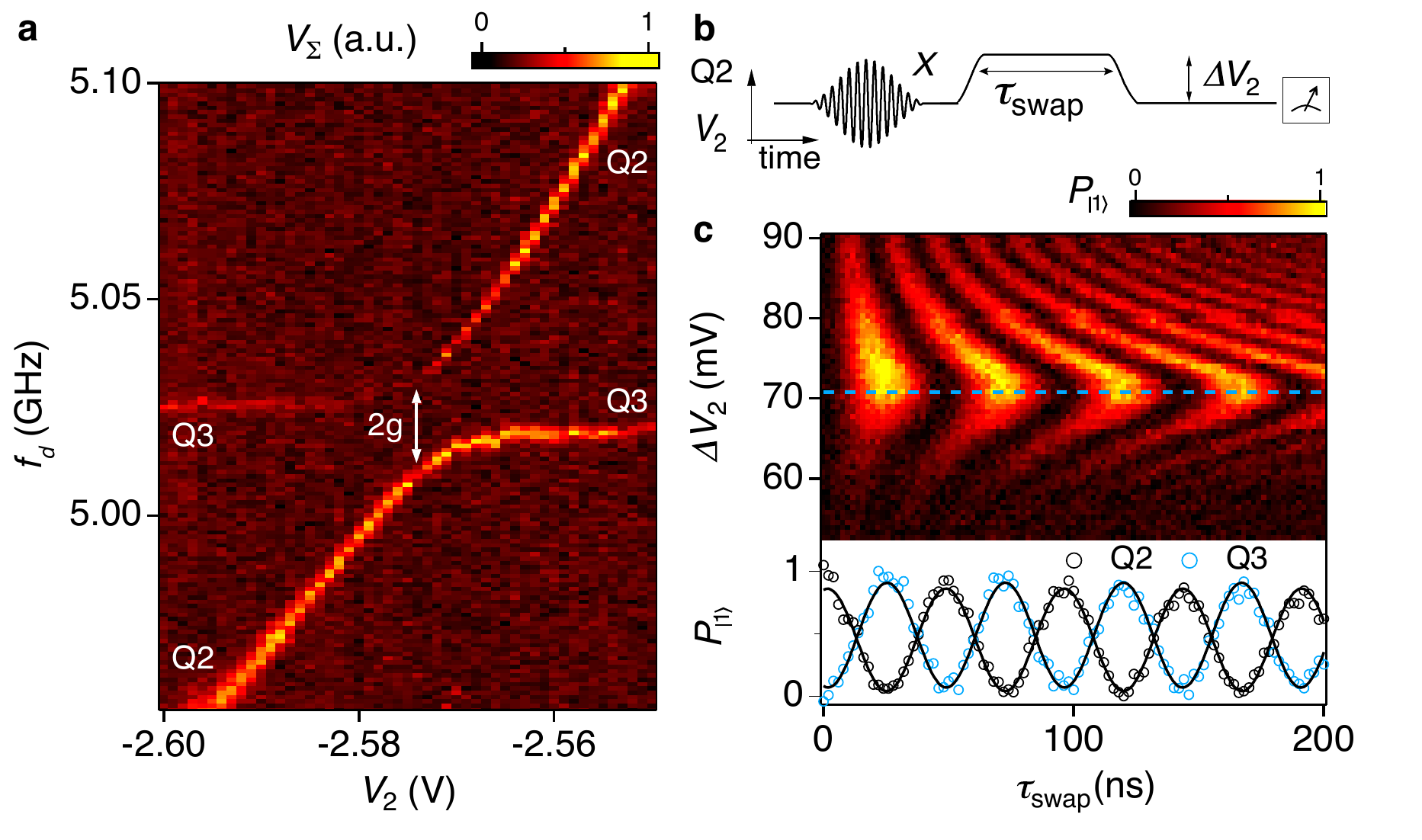}\vspace{-4mm}
    \caption{\textbf{Coherent two-qubit interaction.} \textbf{a,} Measurement of the avoided level crossing between Q2 and Q3. The sum of the normalized heterodyne readout amplitude $V_{\rm{\Sigma}}$ for both qubits is shown as a function qubit drive and $V_2$. \textbf{b,} Pulse sequence to probe the coherent coupling between the qubits. With Q2 and Q3 detuned, Q3 is prepared in the ground state and a $X$ pulse prepares Q2 in \rm{$|1\rangle$}. A gate pulse with amplitude $\Delta V_2$ brings Q2 close to or in resonance with Q3 for time $\tau_{\rm{swap}}$. \textbf{c,} The \rm{$|1\rangle$} state probability, $P_{\rm{|1\rangle}}$, for Q3 as a function of $\Delta V_2$  and $\tau_{\rm{swap}}$. Lower panel: $P_{\rm{|1\rangle}}$  for both Q2 and Q3 at $\Delta V_2\sim$~72~mV, that brings the two qubits into resonance.
    }
    \label{swap}\vspace{-4mm}
\end{figure*}

To measure the relaxation time, $T_1$, an $X$ pulse excites the qubit [blue pulse sequence, Fig.~\ref{times}(a)] and $P_{|1\rangle}$ is plotted as a function of $\tau$, the time delay before readout. The probability decreases exponentially due to relaxation. Fitting the decay (blue) yields $T_1=$~1.1~$\mu$s. To extract the dephasing time $T_2^*$, two slightly detuned $X/2$ pulses are applied, black pulse sequence Fig.~\ref{times}(a) inset, separated by delay time $\tau$. A fit to the decay of the resulting Ramsey fringes [inset Fig.~\ref{times}(a)] gives a dephasing time $T_2^*=$~400~ns. In order to reduce inhomogeneous dephasing due to low-frequency noise, we perform a Hahn echo sequence comprising a refocusing $X$ pulse between two $X/2$ pulses [Fig.~\ref{times}(a),~red]. The extracted $T_{\rm{2,echo}}=$~2.2~$\mu$s~$\sim 2T_1$ indicates that 2DEG gatemon dephasing is dominated by low frequency noise ~\cite{bylander_2011}. Figure~\ref{times}(b) shows $T_1$ as a function of qubit frequency. Relaxation times vary between 0.2 - 2~\rm{$\mu$}s and fluctuate strongly with $f_Q$ [the spectrum is plotted in the inset Fig.~\ref{times}(b)]. Owing to their periodicity, we attribute these fluctuations to on-chip modes, which is consistent with previous results from devices lacking crossover wire bonds. 

An estimate for the dielectric loss of the qubit capacitor can be made using a test resonator coupled to the same feedline ($f_{\rm{res}}$~=~5.35~GHz) which shows an internal quality factor $Q\sim 6.4 \times 10^4$ at low photon number. Using the expression $T_1=Q/(2\pi f_Q)$ we expect the relaxation time due to dielectric loss to follow the black dashed line~\cite{barends_2013}. The agreement between the measured $T_1$ times and this upper bound suggests the qubit lifetime is indeed limited by dielectric loss. Similar $Q$'s are obtained on pure semi-insulating InP substrates, suggesting the presence of the 2DEG does not introduce additional loss. Potential solutions for reducing microwave loss in InP include deep-etching~\cite{bruno_2015} and using flip-chip techniques to host the qubit island on a low-loss substrate such as Si~\cite{rosenberg_2017}. Using the measured lever arm of 1~GHz/V and gate voltage fluctuations of $\sim$~8~$\mu$V measured in III-V semiconductors~\cite{dial_2013}, we expect an upper limit of $T_2^*\sim $~20~$\mu$s, suggesting the prospects are good for obtaining coherence times comparable with state-of-the-art flux-tuneable transmons, where $T_1\sim$~30~$\mu$s and $T_2^*\sim$~5~$\mu$s~\cite{kelly_2015}.


Finally, we coherently swap excitations using the capacitive coupling between neighbouring qubits. The monotonicity and reproducibility of the qubit spectra established in Fig.~\ref{coh} are critical for tuning adjacent qubits into resonance with each other. The signature of qubit-qubit coupling is a mutual push on the bare qubit frequencies due to hybridization. To detect this push, the qubits Q2 and Q3 are driven and read out through the feedline and their respective cavities. For clarity the signals detected from both cavities are added to yield the sum $V_{\Sigma}$. 
Figure~\ref{swap}(a) shows $V_{\Sigma}$ as a function of the qubit drive and $V_2$. As expected, due to the absence of cross-talk, there are two peaks in $V_{\Sigma}$ as a function of $f_d$, only one of which (Q2) is tuned by $V_2$. When tuned onto resonance, the qubits anticross and a splitting of $2g/2\pi\sim$~12~MHz between the two hybridized states is observed, where $g$ is the qubit-qubit coupling strength. Figure ~\ref{swap}(b) shows the pulse sequence that exploits the anticrossing to coherently transfer an excitation between Q2 and Q3, the starting point for preparing arbitrary two-qubit states. With the two qubits detuned by $\sim$~140~MHz and Q3 idling, Q2 is prepared in $|1\rangle$. A gate pulse is then applied for time $\tau_{\rm{swap}}$ and brings Q2 into resonance with Q3~\cite{majer_2007}. Note that here the microwave pulses are applied through the gate line, demonstrating qubit manipulation using a single control line per qubit. We emphasize that single-gate control of rotations around the X, Y and Z-axis is an important advantage of voltage-controlled qubits. The rate at which excitations swap between qubits depends on $\tau_{\rm{swap}}$ and the pulse amplitude $\Delta V_2$. Figure~\ref{swap}(c) shows the typical chevron pattern of swap oscillations~\cite{hofheinz_2009}. The lower inset in Fig.~\ref{swap}(c) shows $P_{|1\rangle}$ for each qubit separately. The anticorrelation confirms that the excitation is transferring between Q2 and Q3 and demonstrates the possibility of generating entangled states using 2DEG gatemon qubits. From sinusoidal fits (solid lines) an interaction rate $2g/2\pi=$~14~MHz is extracted, in good agreement with electrostatic simulations yielding $2g/2\pi \sim$~15~MHz for $f_{Q}=$~5~GHz.

In summary, we have demonstrated that planar semiconductor materials and superconducting microwave circuits are compatible technologies that can be readily integrated while maintaining quantum coherence. This opens new possibilities for highly integrated quantum processors with on-chip components. Through a combination of geometry and applied voltages, $E_J$ can be tailored to simultaneously suit qubits as well as peripheral control circuits requiring higher $E_J$, such as tuneable couplers~\cite{chen_2014, casparis_2018} and on-chip microwave sources \cite{cassidy_2017}, and develop naturally into 3D architectures required to implement fault-tolerant processing~\cite{versluis_2016,rosenberg_2017}. Moreover since 2DEG gatemons represent a perfect quantum counterpart to semiconductor-based cryogenic classical control logic~\cite{ward_2013,al-taie_2013,hornibrook_2014}, they take the first step towards realizing a scalable all-electric hybrid superconductor-semiconductor quantum processor.

\section{Methods}
\subsection{The sample}
Separate transport characterization shows that the 2DEGs exhibit a Hall mobility of approximately 2000~cm\textsuperscript{2}V\textsuperscript{-1}s\textsuperscript{-1} and an induced gap of 200~$\rm{\mu eV}$. 
The qubits were fabricated by first wet etching a mesa for the qubit JJ. The width $w$ of the JJ was defined by the mesa etch. The JJ was then formed by selectively wet etching a $\sim$~100 nm long segment of the $\sim$~50 nm thick Al. A 20~nm thick AlOx layer (yellow in Fig.~\ref{device}(b)-(c)) was deposited as a gate dielectric, followed by the evaporation of an Al top gate (red). The heterostructure and the buffer were removed almost everywhere on the chip, leaving only a few micron large mesa region to form the active region of the qubit. The qubit islands, gate lines and readout cavities were defined in a lift-off process with a 100~nm Al layer. Finally, the epitaxial Al layer on top of the mesa and the microwave circuit were connected in a contact step. For each qubit, $E_C/h$ is determined by the capacitance of the T-shaped Al island to the surrounding ground plane and designed to be $\sim$~230 MHz. 
All qubits are coupled to individual $\lambda$/4 superconducting cavities with resonant frequencies separated by 50~MHz and centred around 7.25~GHz. All six cavities are coupled to a common feed line \cite{barends_2013}.

\subsection{Qubit manipulation and readout}
All measurements presented in the paper are performed in a cryogen-free dilution refrigerator with a base temperature below 50 mK. The sample is mounted inside an Al box to suppress magnetic fluctuations. This box is placed inside a Cu box used to mount the sample at the mixing chamber plate of the refrigerator. Both boxes are closed but not light tight and are further surrounded by a cylindrical cryoperm shield, which is also thermally anchored to the mixing chamber.

The qubit is initialized in the $|0\rangle$ state by waiting for much longer than the relaxation time $T_1$.
To manipulate a single qubit, one coaxial line and a DC line are used: the coax line is filtered by a Minicircuits VLF-320 low pass filter and an ECCOSORB filter to reduce noise while allowing for gate pulses. At high frequencies ($>$2~GHz), the filter attenuates by roughly 20~dB allowing to drive the qubit directly. The DC line is filtered with an RC filter and added with a bias T at low temperature.  For X microwave control as well as readout, the pulses  are shaped through IQ modulation of the microwave source and using an AWG channel for I and Q respectively. For readout, the signal line is heavily attenuated (60~dB) to reduce both the thermal occupation of the resonator and noise to the sample. 
After passing through a magnetically shielded isolator, a travelling wave parametric amplifier~\cite{macklin_2015}, another magnetically shielded isolator, a cryogenic low noise factory HEMT amplifier and another amplification stage at room temperature the qubit readout signals are mixed down to intermediate frequencies with a local oscillator, before sampling and performing digital homodyne detection to extract the cavity magnitude response. Qubit state measurements are obtained by averaging over $\sim$~1000 experimental runs. We use the raw Rabi oscillation data for qubit state assignments. 
The data in Fig.~\ref{coh}(a) were acquired with a vector network analyzer. The data in all other figures were acquired using heterodyne detection in the dispersive regime. For Fig.~\ref{swap} we combine two drives with frequencies close to the resonance frequencies of the cavities of Q2 and Q3 on the signal line. 

\begin{acknowledgments}
\section{Acknowledgments}
We acknowledge helpful discussions with A.~C.~C.~Drachmann, H.~J.~Suominen, E.~O'Farrell, A.~Fornieri, A.~Whiticar and F.~Nichele. This work was supported by Microsoft Project Q, the U.S. Army Research Office, the Innovation Fund Denmark, and the Danish National Research Foundation. C.M.M. acknowledges support from the Villum Foundation. MRC was supported by a Marie Curie Fellowship and EPSRC (EP/L020963/1). MK gratefully acknowledges support from the Carlsberg Foundation. The travelling wave parametric amplifier used in this experiment was provided by MIT Lincoln Laboratory and Irfan Siddiqi Quantum Consulting (ISQC), LLC via sponsorship from the US Government.
\end{acknowledgments}

\section{Author contributions}
T.W., C.T., S.G., G.C.G. and M.J.M. grew the proximitized 2DEG.  M.K., L.C., C.M.M. and  K.D.P designed the experiment. L.C., M.R.C., A.K., N.J.P., T.W.L., and  K.D.P. prepared the experimental setup. L.C. and M.R.C. fabricated the devices and performed the experiment. L.C., M.R.C., M.K., A.K., T.W.L., F.K., C.M.M.,  and  K.D.P. analyzed the data and prepared the manuscript. 

\section{Competing financial interest}
The authors declare no competing financial interests.


\end{document}